\documentclass[12pt]{article}
\usepackage[left=2cm,top=2cm,right=2cm,bottom=2cm]{geometry}

\usepackage[utf8]{inputenc}
\usepackage[british]{babel}

\usepackage{amsfonts,amsmath,amssymb,graphicx,latexsym,xcolor,mathtools}
\usepackage[scr=boondox]{mathalpha}
\usepackage{hyperref,enumitem,enumerate,float}
\usepackage{amsthm,verbatim}
\usepackage{cite}
\usepackage[autostyle]{csquotes}

\numberwithin{equation}{section}
\pdfoutput=1

\theoremstyle{plain}

\theoremstyle{definition}

\newtheorem{defi/}[theorem]{Definition}
\newtheorem{rema/}[theorem]{Remark}
\newtheorem{exa/}[theorem]{Example}

\makeatletter
\newcommand*{\transpose}{%
	{\mathpalette\@transpose{}}%
}
\newcommand*{\@transpose}[2]{%
	\raisebox{\depth}{$\m@th#1\intercal$}%
}
\makeatother

\newcommand{\tensortwo}[1]{\underline{\underline{\boldsymbol{#1}}}}

\newcommand{\vect}[1]{\boldsymbol{#1}}
\newcommand{\tensorind}[2]{\underline{\underline{\boldsymbol{#1}_{#2}}}}

\definecolor{myred}{RGB}{160,0,0}
\definecolor{mygreen}{RGB}{0,160,0}
\definecolor{myblue}{RGB}{0,0,160}
\hypersetup{
	colorlinks = true,
	linkcolor = {myred},
	citecolor = {mygreen},
	urlcolor  = {myblue},
}

\title{Compressional-shear wave interference and wave suppression at exceptional-point gaps}

\author{Gabriel N\'u\~nez, William J. Parnell and Rapha\"{e}l C. Assier\\
	\footnotesize{Department of Mathematics, University of Manchester, Oxford Road, Manchester, {\rm M13 9PL}, UK}
}

\date{}

\begin{document}

\maketitle

\begin{abstract}
Exceptional points (EPs) in complex elastic materials signify the coalescence of compressional and shear modes and bound a frequency gap where waves become evanescent. While their effects on wave propagation have been extensively studied, the physical mechanisms underlying their existence remain elusive. In this Letter, employing a transfer matrix formulation combined with Bloch’s theorem, we demonstrate that the gaps bounded by EPs in layered media originate from destructive interference between the compressional and shear wave fields. We further reveal that the corresponding wavenumbers satisfy a Bragg-like condition with the unit cell thickness, causing the total wave field to become stationary, forming a standing wave. Finally, a finite layered medium is considered, where this mechanism is shown to manifest as forbidden wave propagation for both modes in the gap. These results provide a direct physical interpretation of EPs.
\end{abstract}

\section{Introduction}

Wave propagation in complex elastic media is a well-developed research field \cite{Thomson50,Graf,Brekhovskikh,Whitham,Kinsler,Blackstock,Anselmet,Royer,Wei,Nayfeh91,Amirkhizi17}. 
A wide range of physical phenomena have been investigated in these materials, including surface waves \cite{Camley83,Wei}, band gaps \cite{Camley83,Kushwaha93,Kushwaha94}, anisotropic effective properties \cite{Postma55,Backus62,Nunez25}, and, more recently, the existence of exceptional points (EPs) \cite{Lustig19,Mokhtari20,Elbaz22,Fishman24,Alizadeh25,Nunez26}.

Mathematically, given a parameter-dependent matrix, EPs correspond to a given point in the parameter space for which two or more eigenvalues and their corresponding eigenvectors coalesce \cite{Kato,Heiss90,Heiss00,Dembowski01,Wu04,Seyranian05,Heiss12,Mace12}.
They have been extensively studied in the context of wave propagation in complex media and non-Hermitian systems, including phononic crystals \cite{Christensen16,Lu18}, composite materials \cite{Wu04,Lustig19,Alizadeh25}, waveguides \cite{Xiong16,Xiong17,Ji19,Nennig20,Lawrie22}, and metamaterials \cite{Xiong17,Achilleos17,Wang22}.
EPs are alse commonly found in systems exhibiting parity-time symmetry \cite{Heiss12,Zhu14,Christensen16,Achilleos17,El-Ganainy18,Ji19}, where they can arise through the introduction of balanced loss and gain.
Additionally, when the parameter space is analytically continued into the complex plane, EPs located away from the real axis manifest as avoided crossings (ACs) along the real axis, where two mode branches approach but repel rather than intersect \cite{Heiss00,Dembowski01,Wu04,Seyranian05,Heiss12,Mace12,Lu18,Nunez26}.

More recently, EPs have been reported in passive elastic media, where they arise through the coupling between compressional and shear modes.
Robert et al.\ \cite{Robert06} reported band gaps associated with mode hybridisation in elastic media containing cylindrical inclusions. 
Lustig et al.~\cite{Lustig19} subsequently identified EPs in layered media, notably demonstrating their effects on wave propagation, including negative refraction and beam splitting.
Mokhtari et al.\ \cite{Mokhtari20} investigated the same problem, identifying a new class of EPs and 
%
suggesting the possibility of resonance trapping.
%
%
Elbaz et al. \cite{Elbaz22} demonstrated that encircling EPs in a periodic laminate by perturbating its stiffness induces anomalous scattering.
%
%
%
Fishman et al.\ \cite{Fishman24} reported third-order EPs and demonstrated the existence of axially frozen modes.
Alizadeh and Amirkhizi \cite{Alizadeh25} further 
%
emphasised that EPs occur in pairs and bound gaps (which we refer to as EP gaps) within which the Bloch wavenumbers become complex, giving rise to spatially attenuating Bloch waves.
Finally, N\'u\~nez et al.~\cite{Nunez26} showed that EPs can also occur in shear-modulus space, and that their occurrence depends strongly on the wave propagation angle. 
Despite this extensive research into the effects of EPs on elastic wave propagation, the physical mechanism underlying their formation and that of their associated gaps remains unclear. 
Moreover, wave transmission through layered media within EP gaps has not been investigated in detail.

In this Letter, following previous works \cite{Alizadeh25,Nunez26}, we study wave propagation in a periodic layered medium employing a transfer matrix method \cite{Thomson50,Brekhovskikh,Folds77,Scharnhorst83,Brouard95,Mead96,Allard,Jimenez} combined with Bloch's theorem \cite{Brillouin, Esquivel94,Cao95,Sigalas05,Wen05, Hussein06,Phani06,Hussein09,Srikantha11, Lee15,Alizadeh25}. 
The resulting eigenvalue problem determines the Bloch wavenumber and, consequently, the band structure of the medium. 
We employ this framework to characterise the mathematical nature of the EPs and show that, within the EP gaps, the total field given by the sum of compressional and shear contributions factorises into a short-wavelength component and an envelope function.
Crucially, the wavenumber of the short-wavelength component satisfies a Bragg-like condition involving the unit cell thickness, causing the interference pattern to become spatially stationary and form a standing-wave pattern. 
This reveals that EP gaps originate from destructive interference between compressional and shear contributions to the total field. 
Finally, we consider wave propagation in a finite layered medium within an EP gap, demonstrating that, unlike conventional band gaps, which can suppress one wave mode independently of the other, neither compressional nor shear waves can propagate through the material.

\section{Exceptional points in periodic layered media} \label{Sec_infinite}

\begin{figure}[h!]
    \centering
    \includegraphics[width=0.75\linewidth]{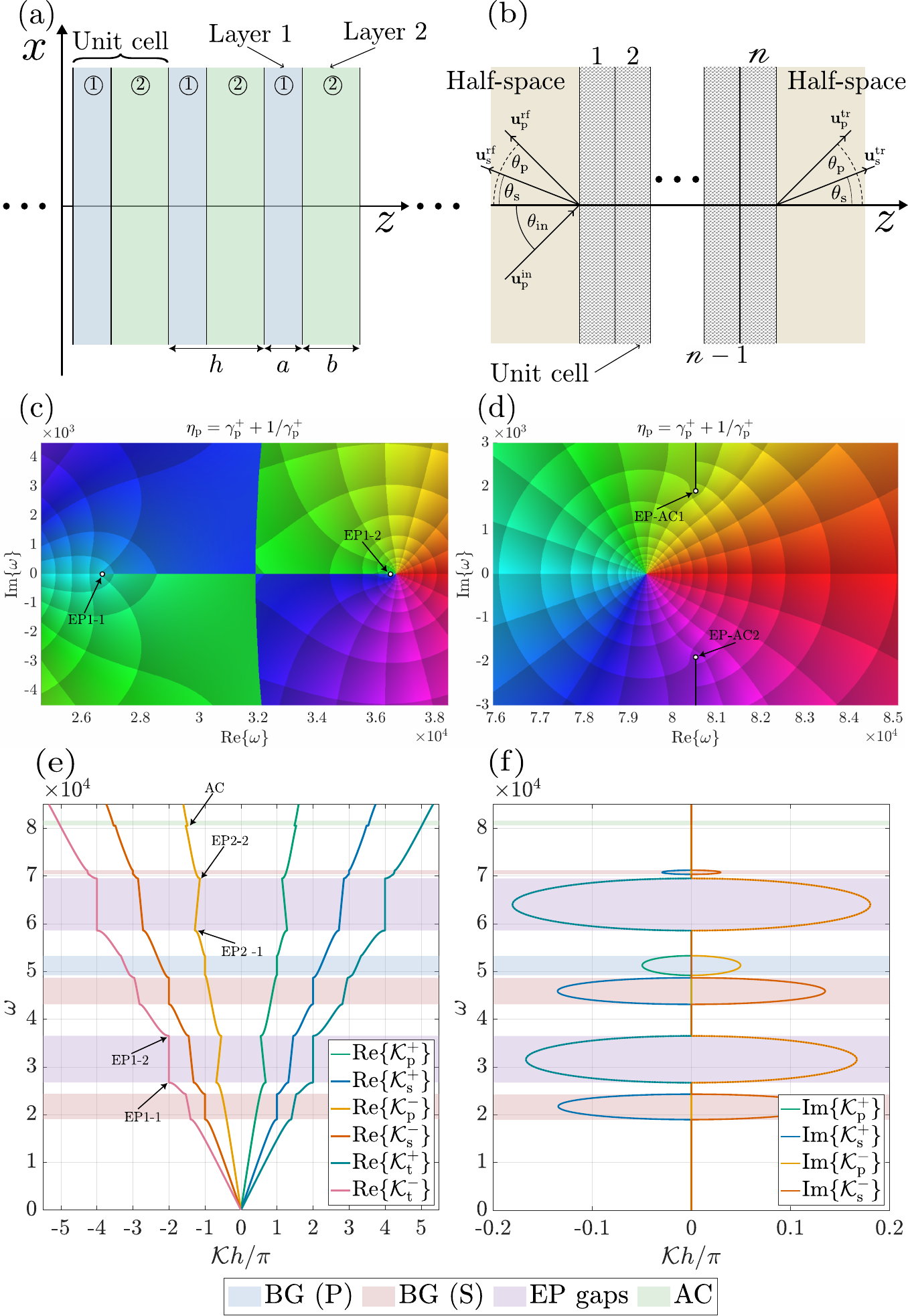}
    \caption{Sketch of the geometry of an infinitely periodic two-layer medium (a) and of a finite two-layer medium with $\mathscr{n}$ unit cells placed between two identical half-spaces (b). Phase portraits of $\eta_\mathrm{p} = \gamma_\mathrm{p}^+ + 1/\gamma_\mathrm{p}^+$ around an EP gap (c) and an AC (d). Dispersion diagram showing the real (e) and imaginary (f) parts of the Bloch wavenumber for compressional and shear waves as functions of frequency, where $\mathcal{K}_t^\pm = \mathcal{K}_\mathrm{p}^\pm + \mathcal{K}_\mathrm{s}^\pm$.  The integers $\mathscr{k}_\mathrm{m}$ in \eqref{Eq_Bloch_wn_gen} are chosen such that $\mathrm{Re}\{\mathcal{K}_\mathrm{m}\}$ varies continuously in frequency.}
    \label{Fig_1}
\end{figure}

Consider an infinite periodic material, shown in Figure \ref{Fig_1}a, whose unit cell consists of two homogeneous isotropic elastic layers with thicknesses $a$ and $b = h - a$, where $h$ is the unit cell thickness.
The layers, labelled Layer 1 and Layer 2, have densities $\rho_\ell$,  Young's moduli $E_\ell$, and Poisson's ratios $\nu_\ell$, with $\ell \in \{1,2\}$ indicating the layer number.
Without loss of generality, under the assumption of plane strain, the problem can be considered in the $xz$-plane, with the $x$-axis parallel to the layer boundaries and the $z$-axis perpendicular to them. 
The layers admit time-harmonic compressional ($\mathrm{p}$) and shear ($\mathrm{s}$) wave propagation, with angular frequency $\omega$ and propagation angles $\theta_{\mathrm{m}_{\ell}}$ with respect to the $z$-axis, with $\mathrm{m}\in\{\mathrm{p},\mathrm{s}\}$.

Wave propagation in the unit cell is described via its non-Hermitian transfer matrix $\tensortwo{T}$.
Its derivation is given in Appendix \ref{Append_A}.
To model the periodic medium, $\tensortwo{T}$ is combined with Bloch's theorem.
In this setting, Bloch's theorem states
\begin{equation} \label{Eq_bloch_theo}
    \vect{V}(x,h) = \exp{(i \mathcal{K}h)} \vect{V}(x,0),
\end{equation}
where the state vector $\vect{V}(x,z)$ is given in \eqref{Eq_transfer_matrix_state}.
Combining \eqref{Eq_bloch_theo} with \eqref{Eq_acous_tm_layer_rel} gives
\begin{equation} \label{Eq_system_eigs}
    \left( \tensortwo{T} - \gamma_\mathrm{m}\tensortwo{I}\right) \vect{V}(x,0) = \vect{0},
\end{equation}
where $\gamma_\mathrm{m}$ are the eigenvalues of $\tensortwo{T}$ and the Bloch wavenumber is $\mathcal{K}_\mathrm{m} = - i \ln{(\gamma_\mathrm{m})}/h$.
For complex $\gamma_\mathrm{m}$, the natural logarithm is defined as ${\ln{\gamma_\mathrm{m}} \coloneq \ln{|\gamma_\mathrm{m}|} + i \left( \varphi_{\gamma_\mathrm{m}} + 2 \pi \mathscr{k}_{\mathrm{m}} \right)}$, with $\varphi_{\gamma_\mathrm{m}}$ denoting the phase of $\gamma_\mathrm{m}$ and $\mathscr{k}_{\mathrm{m}} \in \mathbb{Z}$, leading to
\begin{equation} \label{Eq_Bloch_wn_gen}
    \mathrm{Re}\{\mathcal{K}_\mathrm{m}\} = \frac{\varphi_{\gamma_\mathrm{m}} + 2 \pi \mathscr{k}_{\mathrm{m}}}{h}, \quad \mathrm{Im}\{\mathcal{K}_\mathrm{m}\}\ = - i \frac{\ln{|\gamma_\mathrm{m}|}}{h}.
\end{equation}

The eigenvalue problem in \eqref{Eq_system_eigs} leads to two compressional and two shear eigenvalues, labelled $\gamma_\mathrm{p}^\pm$ and $\gamma_\mathrm{s}^\pm$, respectively.
They satisfy $\gamma_\mathrm{m}^- = 1/\gamma_\mathrm{m}^+$ and are grouped via the auxiliary variable $\eta_\mathrm{m} = \gamma_\mathrm{m}^+ + 1/\gamma_\mathrm{m}^+$.
They are calculated from \eqref{Eq_tm_eigs_gamma}--\eqref{Eq_tm_eigs_eta}.
The forward branch (positive $z$-direction) is identified with $^+$ while the backward branch (negative $z$-direction) with $^-$.
Additionally, this allows to define the eigenvectors as $\vect{V}_\mathrm{m}^\pm$, obtained by solving \eqref{Eq_system_eigs} for the respective eigenvalue.
The EPs occur when the eigenvalues and their corresponding eigenvectors for the two modes coalesce.
This occurs when the two $\eta_\mathrm{m}$ branches merge, leading to $\gamma_\mathrm{p}^+ = \gamma_\mathrm{s}^+$ and $\gamma_\mathrm{p}^- = \gamma_\mathrm{s}^-$.
From \eqref{Eq_tm_eigs_eta}, this demonstrates that EPs correspond to branch points of the eigenvalues as functions of $\omega$, agreeing with the results of \cite{Elbaz22}.
Importantly, EPs only occur under oblique incidence.
For normal incidence, the wave modes are decoupled and the eigenvectors remain linearly independent at the coalescing points. 
These are called \textit{diabolical points} \cite{Seyranian05}.

Since the EP gaps occur due to the coalescence of the compressional and shear eigenvalues, their existence must involve an interaction between these two wave types.
In a conventional band gap, wave propagation is forbidden because of Bragg scattering \cite{Jimenez}.
This occurs when $\mathcal{K}_\mathrm{m} = n \pi/h$ for $n \in \mathbb{N}$.
To derive an analogous expression for the EP gap, we focus on the state vector $\vect{V}_\mathrm{m}(x,z)$ associated with $\gamma_{\mathrm{m}}$.
From now own, we omit the superscript $^\pm$ for simplicity.
Let the field ${v_\mathrm{m}(x,z) \in \vect{V}_\mathrm{m}(x,z)}$ be a component of the state vector. 
Eq.~\eqref{Eq_bloch_theo} then becomes ${v_\mathrm{m} (x,z) = v_{\mathrm{m}_0} \exp{(i \mathcal{K}_\mathrm{m} z)} \exp{(i k_x x)}}$, where ${v_{\mathrm{m}_0} = v_\mathrm{m}(x,0)}$ and $k_x$ is the wavenumber in the $x$-direction.
Since, due to Snell's law, $k_x$ is conserved across the layers, the $x$-dependence is now omitted for ease of notation.
The total field $v_t$ is the sum of the compressional and shear components, i.e., 
\begin{equation} \label{Eq_sum_Bloch_fields}
    v_t(z) = v_{\mathrm{p}_0} \exp{(i \mathcal{K}_\mathrm{p} z)} + v_{\mathrm{s}_0} \exp{(i \mathcal{K}_\mathrm{s} z)}.
\end{equation}
We express the field as ${v_{\mathrm{m}_0} = |v_{\mathrm{m}_0}| \exp{(i \varphi_{\mathrm{m}_0})}}$, so that $\varphi_{{\mathrm{m}_0}}$ denotes the phase of ${v_{\mathrm{m}_0}}$.
From Appendix \ref{Append_B}, it follows that, within the EP gap, ${|v_{\mathrm{p}_0}| = |v_{\mathrm{s}_0}| \coloneq |v_{0}|}$ and either ${\varphi_{{\mathrm{p}_0}} = - \varphi_{{\mathrm{s}_0}} \coloneq \varphi_{0}}$ or ${\varphi_{{\mathrm{p}_0}} = \pi - \varphi_{{\mathrm{s}_0}} \coloneq \varphi_{1}}$, depending on the choice of normalisation. 
Choosing the former, for \eqref{Eq_sum_Bloch_fields} this entails that
\begin{equation} \label{Eq_total_field_ph}
    v_{t}(z) = |v_{0}| \left[ \exp{(i \mathcal{K}_\mathrm{p} z + i \varphi_{0})} + \exp{(i \mathcal{K}_\mathrm{s} z - i \varphi_{0})} \right].
\end{equation}
Taking the real part of \eqref{Eq_total_field_ph} and applying the angle-addition identities for the trigonometric functions allows us to write ${\mathrm{Re}\{v_{t}(z)\} = 2 |v_{0}| S(z) L(z)}$, with
\begin{align}
    S(z) &= \cos{\left( \frac{\mathrm{Re}\{\mathcal{K}_\mathrm{p} + \mathcal{K}_\mathrm{s}\}}{2} z \right)}, \label{Eq_short_factor} \\ 
    L(z) &= \cos{\left( \frac{\mathrm{Re}\{\mathcal{K}_\mathrm{s} - \mathcal{K}_\mathrm{p}\}}{2} z - \varphi_0\right)}  \exp{(-\mathcal{K}_v z)}, 
\end{align}
and where ${|\mathcal{K}_v| \coloneq |\mathrm{Im}\{\mathcal{K}_\mathrm{p}\}| = |\mathrm{Im}\{\mathcal{K}_\mathrm{s}\}|}$, and the exponential decay is determined by $\exp{(-\mathcal{K}_v z)}$.
Choosing $\varphi_{1}$ instead of $\varphi_{0}$ yields analogous expressions.
This result indicates that the total field can be expressed as the product between a \textit{short}-wavelength factor $S(z)$ and an \textit{envelope} function $L(z)$.
Crucially, within the EP gap it holds that
\begin{equation} \label{Eq_sum_wavenumbers_ep}
    \mathrm{Re}\{\mathcal{K}_t\} = \mathrm{Re}\{\mathcal{K}_\mathrm{p} + \mathcal{K}_\mathrm{s}\} = \frac{2\pi n}{h} \ \text{for} \ n \in \mathbb{N}.
\end{equation}
This result is demonstrated in Appendix \ref{Append_B}.
Consequently, the wavelength of $S(z)$ is
\begin{equation} \label{Eq_sum_wavelengths_ep}
    \lambda_S = \frac{4 \pi }{\mathrm{Re}\{\mathcal{K}_\mathrm{p} + \mathcal{K}_\mathrm{s}\}} = \frac{2h}{n}.
\end{equation}
This implies that an integer number of half-wavelengths of $S(z)$ fits within the unit cell. 
This condition is analogous to the Bragg condition for conventional band gaps, but involves the sum of the compressional and shear wavenumbers rather than a single mode.
Moreover, the short-wavelength component forms a standing wave pattern within the unit cell.
Consequently, we propose the following physical interpretation:

\indent \textit{EP gaps arise when the superposition of compressional and shear waves produces destructive interference in the total wave field, whose short-wavelength component satisfies a Bragg-like condition with the unit cell thickness and forms a standing wave pattern within the unit cell.}

To further demonstrate the behaviour of the EPs, we investigate a specific configuration.
In this example, Layer 1 consists of a glass-fibre-reinforced polymer (GFRP) \cite{Shakir21,Gribniak21}, while Layer 2 consists of glass-ceramic AR 3 \cite{Suputtamongkol22}.
Their physical properties are given in Table \ref{Table_mat_properties}, together with polycarbonate \cite{Alizadeh25} which is employed in a later example.
The layer thicknesses are $a=0.14\mathrm{m}$ and $b = 0.26\mathrm{m}$.
\begin{table}[h!]
\caption{\label{Table_mat_properties}%
Physical properties of polycarbonate, GFRP and glass ceramic.
}
\centering
\begin{tabular}{|c|c|c|c|}
\hline
\textrm{Material}&
\textrm{$\rho$}&
\textrm{$E$}&
\textrm{$\nu$}\\
 & kg/m$^3$ & GPa & -\\
\hline
Polycarbonate & 1200 & 2 & 0.37\\
GFRP & 1800 & 40 & 0.36\\
Glass ceramic (AR 3) & 4000 & 75 & 0.28\\
\hline
\end{tabular}
\end{table}

The phase portraits of $\eta_\mathrm{p}$ in complex frequency and corresponding dispersion diagrams are presented in Figures \ref{Fig_1}c-\ref{Fig_1}f.
These reveal several of the features of EPs described previously.
Figure \ref{Fig_1}c shows the phase portrait around the first EP pair, labelled EP1-1 and EP1-2 in Figure \ref{Fig_1}e.
As expected, the EPs occur in pairs on the real axis and bound an EP gap.
Figure \ref{Fig_1}d shows the phase portrait around the AC highlighted in green in Figure \ref{Fig_1}e.
This demonstrates that the AC on the real-frequency axis is associated with a pair of complex-conjugate EPs in the complex-frequency plane.

\begin{figure}[h!]
    \centering
    \includegraphics[width=0.75\linewidth]{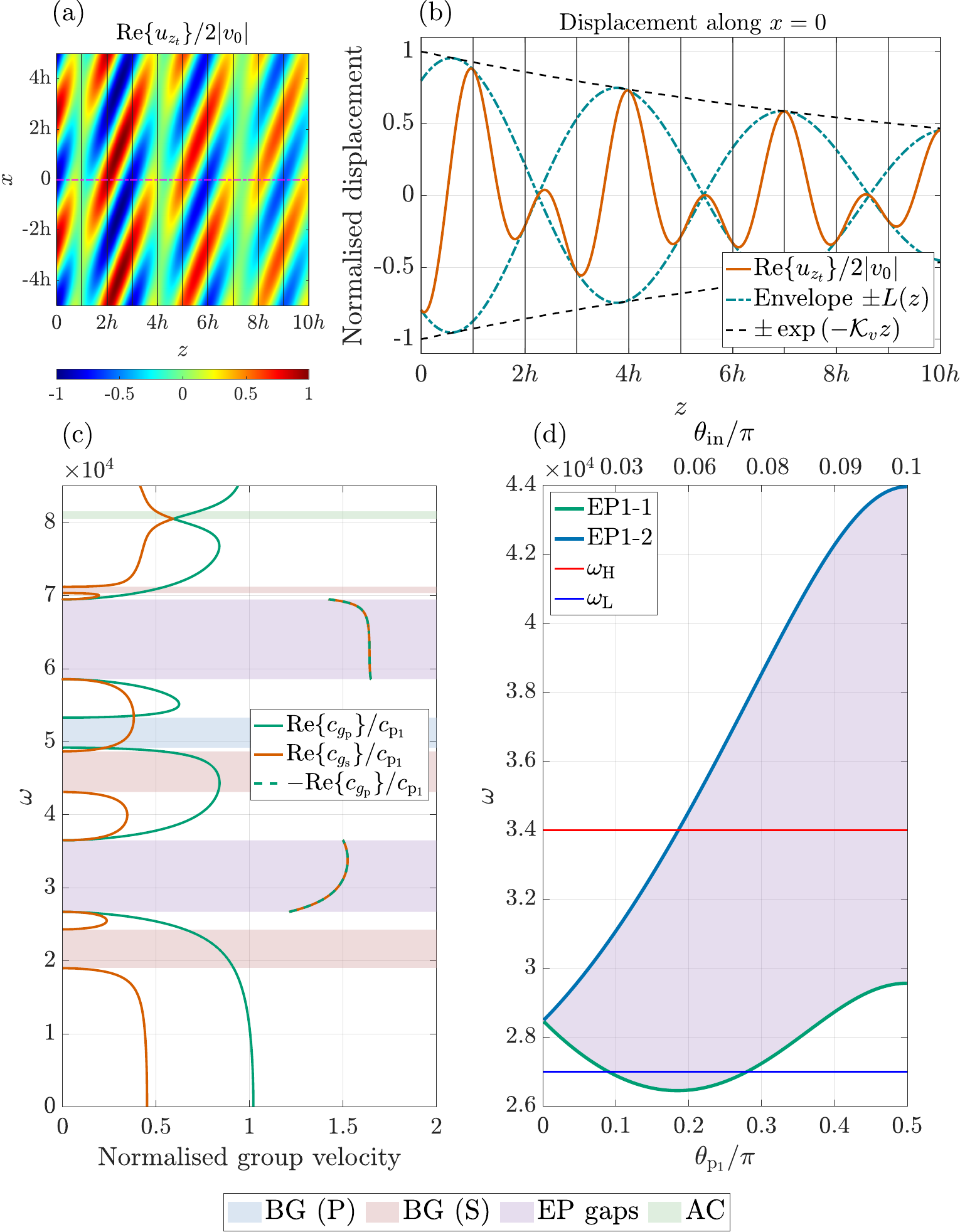}
    \caption{(a) Real part of total displacement field $u_{z_t}$ for a frequency inside an EP gap. (b) Normalised displacement $\mathrm{Re}\{u_{z_t}\}$  for wave propagation along $x = 0$. (c) Compressional and shear group velocities $c_{g_\mathrm{p}}$ and $c_{g_\mathrm{s}}$ as functions of frequency. (d) Frequency trajectories of EP1-1 and EP1-2 as functions of $\theta_{\mathrm{p}_1}$ and $\theta_\mathrm{in}$. The horizontal lines mark the frequencies $\omega_\mathrm{L}$ and $\omega_\mathrm{H}$ for Figure \ref{Fig_3}.}
    \label{Fig_2}
\end{figure}

Figures \ref{Fig_1}e and \ref{Fig_1}f present the dispersion diagrams for both wave modes under oblique wave propagation of $\theta_{\mathrm{p}_1} = \pi/4$ while the remaining angles are calculated via Snell's law as given in \eqref{Eq_snells_law}.
Within the EP gap, the rightmost and leftmost curves showing $\mathrm{Re}\{\mathcal{K}_t\}$ for forward and backward branches, respectively, illustrate the result of \eqref{Eq_sum_wavenumbers_ep}.
Moreover, the wavenumber has a non-zero imaginary part in this range, indicating that both waves are evanescent.
Thus, the EP gap corresponds to a frequency interval in which neither compressional nor shear waves can propagate through the periodic medium.
This contrasts with a conventional band gap, in which only one mode is suppressed while the other can propagate.
Another key distinction between band gaps and EP gaps is that $\mathrm{Re}\{\mathcal{K}_\mathrm{m}\}$ is not constant within an EP gap. 
To illustrate the standing-wave pattern formed within the unit cell, Figure~\ref{Fig_2}a shows the real part of the total displacement field in the $z$-direction, i.e., ${\mathrm{Re}\{u_{z_t}\} = \mathrm{Re}\{u_{z_\mathrm{p}} + u_{z_\mathrm{s}}\}}$, within the first EP gap, meaning $n = 1$ in \eqref{Eq_sum_wavenumbers_ep}.
%
%
The same analysis applies to any component $v_\mathrm{m} \in \vect{V}_\mathrm{m}$.
We then consider the total displacement along $x = 0$, plotted in Figure \ref{Fig_2}b together with the envelope $L(z)$.
The curve $\mathrm{Re}\{u_{z_t}\}$ is bounded by $L(z)$ and touches it whenever $z$ is an integer multiple of $h$.
This behaviour follows from \eqref{Eq_short_factor} and \eqref{Eq_sum_wavenumbers_ep}.
Setting $z=\mathscr{k}h$, where $\mathscr{k} \in \mathbb{Z}$, leads to $|S(z)| = 1$. 
Consequently, the maxima and minima of $S(z)$ align with the unit cell thickness. 
By contrast, the crossings between $\mathrm{Re}\{u_{t_z}\}$ and $L(z)$ at points where $L(z) = 0$ are not associated with the Bragg-like condition.

Further insight into the physical behaviour of the EP gaps can be obtained by examining the group velocities $c_{g_\mathrm{m}} = d \omega / d( \mathrm{Re}\{\mathcal{K}_\mathrm{m}\})$, defined here from the real part of the Bloch wavenumber.
The resulting group velocities are shown in Figure \ref{Fig_2}e.
At the edges of the compressional band gaps, $c_{g_\mathrm{p}}$ tends to zero, while $c_{g_\mathrm{s}}$ tends to zero at the edges of the shear band gaps.
In contrast, at each EP, both group velocities tend to zero. 
Notably, within the EP gap, the dispersion diagram for compressional waves exhibits a negative gradient and $c_{g_\mathrm{p}}$ becomes negative.
This is consistent with anomalous energy transport and negative refraction reported for EP gaps \cite{Lustig19}.

The position of the EPs in the $\omega$-plane also depends on the propagation angles.
Figure \ref{Fig_2}f shows the frequency trajectories of EP1-1 and EP1-2 as functions of ${\theta_{\mathrm{p}_1} \in [0,\pi/2)}$.
The strong dependence of the EP frequencies on the propagation angle is demonstrated.
Importantly, as $\theta_{\mathrm{p}_1} \rightarrow 0$, the EP gap closes as the EPs merge into a diabolical point, consistent with the fact that EPs occur only at oblique incidence.

\section{Finite layered medium} \label{Sec_finite}

We now focus on the case of a finite layered medium subject to an incident compressional plane wave at a frequency that lies within an EP gap.  
%
%
%
%
For this purpose, we consider the unit cell introduced in the previous section.
A total of 20 unit cells are placed between two isotropic homogeneous half-spaces with identical material properties, as shown in Figure \ref{Fig_1}b.
An incident compressional waves impinges on the finite medium from the left half-space with angle $\theta_{\mathrm{in}}$, which is related to $\theta_{\mathrm{p}_1}$ via \eqref{Eq_snells_law}.
Nevertheless, the same conclusions can be reached when considering an incident shear wave.
The half-spaces consist of polycarbonate, whose properties are listed in Table \ref{Table_mat_properties}.

Figures \ref{Fig_3}a and \ref{Fig_3}b show the reflected and transmitted energy coefficients, defined in Appendix \ref{Append_C}, for both waves as functions of $\theta_{\mathrm{in}}$ at frequencies $\omega_\mathrm{L}$ and $\omega_\mathrm{H}$.
These frequencies are chosen so that, as $\theta_{\mathrm{in}}$ varies, the system both enters and exits an EP gap, as shown in Figure \ref{Fig_2}b where $\theta_{\mathrm{in}}$ is displayed in the upper axis.
Importantly, $\theta_{\mathrm{in}} \in \{0,\pi/10)$ since total internal reflection \cite{Blackstock} occurs for $\theta_{\mathrm{in}} \geq \pi/10$.
%
%
In both cases, as $\theta_{\mathrm{in}}$ varies, the transmitted energy rapidly decreases to zero as the system enters an EP gap and increases again as it exits the gap.
The transmitted energy coefficients mostly vanish within the EP gap, i.e., ${\mathcal{E}_\mathrm{p}^\mathrm{tr} = \mathcal{E}_\mathrm{s}^\mathrm{tr} = 0}$.
Thus, the incident compressional wave is completely reflected as a sum of compressional and shear waves.
Conversely, outside the EP gap, the incident energy is distributed between reflected and transmitted waves.
These results demonstrate that no energy is transmitted through the finite medium within the EP gap.

Lastly, Figures \ref{Fig_3}d and \ref{Fig_3}e show the displacement fields inside the finite layered medium for an incident wave with frequencies lying within an EP gap and a compressional band gap, respectively.
Additionally, wave propagation in a pass band is presented in Figure \ref{Fig_3}c for comparison, where both waves can propagate through the medium. 
Within the EP gap, the incident wave is almost completely reflected, and the transmitted field in the right half-space is negligible for both modes.
Consequently, all the displacement fields inside the layered medium become evanescent.
In contrast, in a conventional compressional band gap, the compressional wave is mostly reflected while the shear wave can still propagate in the material.
Therefore, unless the compressional and shear band gaps overlap, only one wave mode becomes evanescent.
Thus, EP gaps provide a mechanism for simultaneously suppressing the propagation of both compressional and shear waves in layered media.

\begin{figure*}[h!]
    \centering
    \includegraphics[width=\linewidth]{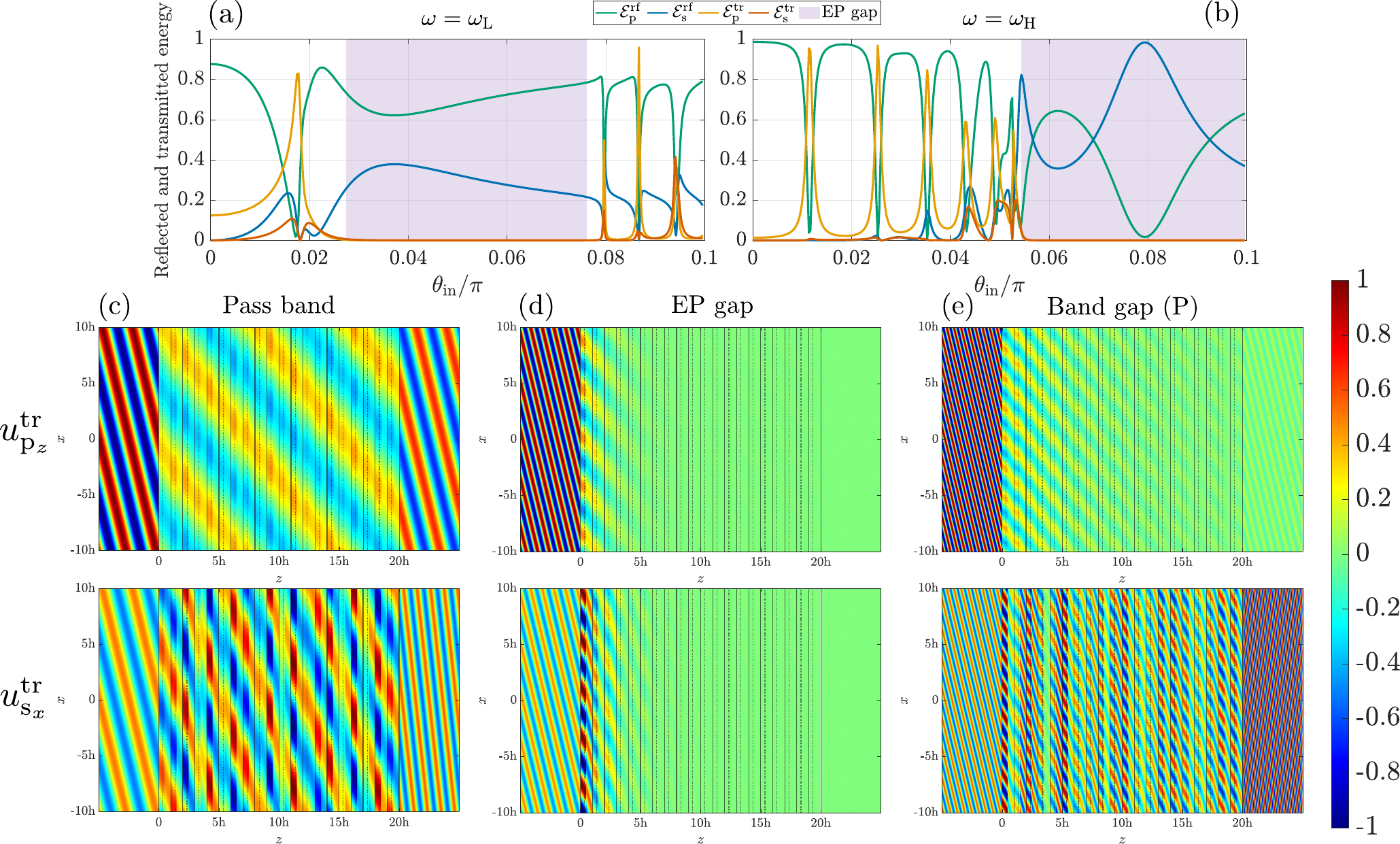}
    \caption{Reflected and transmitted energy coefficients $\mathcal{E}$ for compressional and shear waves for the medium in Figure \ref{Fig_1}b at $\omega_\mathrm{L}$ (a) and $\omega_\mathrm{H}$ (b). Displacement fields of the transmitted compressional and shear waves, represented by $u_{\mathrm{p}_z}^\mathrm{tr}$ and $u_{\mathrm{s}_x}^\mathrm{tr}$, respectively, for an incident compressional wave at a frequency lying within a pass band (c), an EP gap (d) and a compressional band gap (e). The incident compressional field in the left half-space is shown in all cases. All wave fields are normalised to the range $[-1,1]$.}
    \label{Fig_3}
\end{figure*}
\section{Conclusions}

This letter investigated EPs in lossless elastic layered media.
These points occur when a pair of eigenvalues and their corresponding eigenvectors of the non-Hermitian transfer matrix coalesce, and bound an EP gap where both compressional and shear wave modes cannot propagate.
We found that this phenomenon arises from destructive interference between the compressional and shear contributions to the total field.
In the periodic medium, the short-wavelength component of their sum satisfies a Bragg-like condition, resulting in a standing-wave pattern within the medium.

The behaviour of elastic waves in a finite layered medium within an EP gap was also investigated, showing that, in contrast to a conventional band gap, both compressional and shear waves become evanescent.
This result suggests potential applications of EP gaps in wave control, where simultaneous suppression of compressional and shear wave propagation is desirable.

Importantly, the position and extent of the EP gaps depend strongly on the angle of incidence. 
This suggests that reconfigurable layered media, in which the layer orientation can be adjusted based on the incident wave, may provide a route towards wave suppression based on EP gaps, or even act as an ``angular filter'' for incoming waves.

\section*{Funding}

G.N. is supported by the Dean’s Doctoral Scholarship Award of The University of Manchester.
\section*{Appendices}
\setcounter{subsection}{0}
\renewcommand\thesubsection{\Alph{subsection}}
\subsection{Derivation of transfer matrix and eigenvalues} 
\label{Append_A}

\renewcommand{\theequation}{A.\arabic{equation}}

This appendix summarises the derivation of the transfer matrix for the periodic layered medium.
Further details can be found in \cite{Nunez26}.
Within layer $\ell$, the displacements $\vect{u}_{\ell} = (u_{x_{\ell}}, u_{z_{\ell}})$ and stress fields $\tensorind{\sigma}{\ell}$ are gathered into the state vector $\vect{V}_\ell(x,z)$ as
\begin{equation} \label{Eq_transfer_matrix_state}
    \vect{V}_{\ell}(x,z) = \begin{pmatrix}
    u_{x_{\ell}}(x,z) \\
    u_{z_{\ell}}(x,z) \\
    \sigma_{zz_{\ell}}(x,z) \\
    \sigma_{xz_{\ell}}(x,z)
    \end{pmatrix} = \tensorind{X}{\ell} \ \tensorind{D}{z_{\ell}}(z) \ \tensorind{D}{x_{\ell}}(x) \ \vect{A}_{\ell},
\end{equation}
where $\tensorind{X}{\ell}$ is a matrix containing the wavenumbers, polarisation vectors, and elastic constants, $\vect{A}_{\ell}$ is a vector collecting the amplitudes of the forward- and backward-propagating compressional and shear waves, and $\tensorind{D}{z_{\ell}}(z)$ and $\tensorind{D}{x_{\ell}}(x)$ are diagonal matrices containing the exponential terms in $z$ and $x$, respectively.
%

%
The state vectors at $z = 0$ and $z = L$, with $L$ representing the thickness of the layer, are related by
\begin{equation} \label{Eq_acous_tm_layer_rel}
    \vect{V}_{\ell}(x,L) = \tensorind{T}{\ell} \vect{V}_{\ell}(x,0),
\end{equation}
where the transfer matrix for the layer is given by ${\tensorind{T}{\ell} = \tensorind{X}{\ell} \ \tensorind{D}{z_{\ell}}(L) \tensorind{X}{\ell}^{-1}}$.
%
%

%
At each interface, continuity of displacement and traction ${\vect{t} = \tensortwo{\sigma} \cdot \vect{n}}$ are imposed.
The total transfer matrix for $\mathscr{n}$ two-layer unit cells is therefore found as
\begin{equation} \label{Eq_acous_total_tm_n_uc}
    \tensortwo{T}^{(\mathscr{n})} = \left(\tensorind{T}{1} \ \tensorind{T}{2}\right)^\mathscr{n}.
\end{equation}
The transfer matrix of the elastic layered medium is symplectic \cite{Mead96}.
Consequently, its four eigenvalues form two reciprocal pairs $\gamma_\mathrm{p}^- = 1/\gamma_\mathrm{p}^+$ and $\gamma_\mathrm{s}^- = 1/\gamma_\mathrm{s}^+$, and ${\det{(\tensortwo{T})} = 1}$.
The characteristic polynomial therefore can be written as ${P(\gamma_\mathrm{m}) = \gamma_\mathrm{m}^4 - \mathscr{c}_1 \gamma_\mathrm{m}^3 + \mathscr{c}_2 \gamma_\mathrm{m}^2 - \mathscr{c}_1 \gamma_\mathrm{m} + 1}$, where $\mathscr{c}_1$ is the trace of $\tensortwo{T}$ and $\mathscr{c}_2$ is the sum of the determinants of the $2\times2$ principal submatrices of $\tensortwo{T}$, i.e.,
\begin{equation} \label{Eq_constants_c1_c2}
    \mathscr{c}_1 = \mathrm{Tr}(\tensortwo{T}), \quad \mathscr{c}_2 = \frac{1}{2} \left( \left(\mathrm{Tr}(\tensortwo{T})\right)^2 - \mathrm{Tr}(\tensortwo{T}^2) \right). 
\end{equation}
Introducing the auxiliary variable ${\eta_\mathrm{m} = \gamma_\mathrm{m} + 1/\gamma_\mathrm{m}}$ reduces the quartic characteristic equation to ${\eta_\mathrm{m}^2 - \mathscr{c}_1 \eta_\mathrm{m} + \mathscr{c}_2 - 2 = 0}$.
The roots for $\eta_\mathrm{m}$, and consequently the eigenvalues of $\tensortwo{T}$, are given by
\begin{align} \label{Eq_tm_eigs_gamma}
    \gamma_\mathrm{m}^+ &= \frac{\eta_\mathrm{m} + i \sqrt{4 - \eta_\mathrm{m}^2}}{2}, \quad \gamma_\mathrm{m}^- = \frac{\eta_\mathrm{m} - i \sqrt{4-\eta_\mathrm{m}^2}}{2}, \\
    \eta_\mathrm{m} &= \frac{\mathscr{c}_1 \pm \sqrt{\mathscr{c}_1^2 - 4(\mathscr{c}_2 - 2)}}{2}. \label{Eq_tm_eigs_eta}
\end{align}
Eq.~\eqref{Eq_tm_eigs_eta} yields one solution for each wave mode. 
However, the assignment of each solution to a mode depends on the corresponding eigenvectors.
Specifically, eigenvectors satisfying $|u_z| > |u_x|$ and $|\sigma_{zz}|>|\sigma_{xz}|$ are associated with compressional waves, whereas the opposite case indicates shear waves.

From \eqref{Eq_tm_eigs_gamma}, conventional band gaps occur when ${|\gamma_\mathrm{m}|\neq1}$, with their edges corresponding to $\eta_\mathrm{m}^2 = 4$, and the two eigenvalues in each reciprocal pair coalesce at $\gamma_\mathrm{m}=\pm1$.
Within the band gap, the eigenvalues are real, and thus the corresponding Bloch wavenumber is complex with constant real part.
On the other hand, EPs occur when the discriminant of the quadratic equation for $\eta_\mathrm{m}$ vanishes, i.e., when $\mathscr{c}_1^2 = 4(\mathscr{c}_2 - 2)$.

\subsection{Eigenvalues, Bloch wavenumbers, and eigenvectors within an EP gap} \label{Append_B}
\setcounter{equation}{0}
\renewcommand{\theequation}{B.\arabic{equation}}

To derive \eqref{Eq_sum_wavenumbers_ep}, we examine the behaviour of $\eta_\mathrm{m}$ and $\gamma_\mathrm{m}$ within the EP gap.
From the preceding section, ${\mathscr{c}_1^2 < 4(\mathscr{c}_2 - 2)}$ within the gap, implying that ${\eta_\mathrm{m} \in \mathbb{C}}$.
We write ${\eta_\mathrm{m} = |\eta_\mathrm{m}| \exp{i \varphi_{\eta_\mathrm{m}}}}$, where \eqref{Eq_tm_eigs_eta} implies that ${|\eta_\mathrm{p}| = |\eta_\mathrm{s}| \coloneq |\eta|}$ and ${\varphi_{\eta_\mathrm{p}} = -\varphi_{\eta_\mathrm{s}} \coloneq \varphi_\eta}$.
Substituting these representations into \eqref{Eq_tm_eigs_gamma} gives
\begin{equation} \label{Eq_eigs_within_EP}
    \gamma_\mathrm{p}^\pm = \frac{|\eta| e^{i \varphi_\eta} \pm i |A| e^{i \varphi_A}}{2}, \ \gamma_\mathrm{s}^\pm = \frac{|\eta| e^{-i \varphi_\eta} \mp i |A| e^{-i \varphi_A}}{2},
\end{equation}
where $|A|$ and $\varphi_A$ denote the modulus and phase of $\sqrt{4 - \eta_\mathrm{p}^2}$.
Eq.~\eqref{Eq_eigs_within_EP} shows that the compressional and shear eigenvalues, for waves propagating in the same direction, are complex conjugates, so that ${|\gamma_\mathrm{s}^\pm| = |\gamma_\mathrm{p}^\pm|}$ and ${\varphi_{\gamma_\mathrm{s}}^\pm = -\varphi_{\gamma_\mathrm{p}}^\pm}$.
From the definition of the Bloch wavenumber given in \eqref{Eq_Bloch_wn_gen}, this result yields
\begin{equation}
    \mathrm{Re}\{ \mathcal{K}_t^\pm \} = \frac{2 \pi (\mathscr{k}_\mathrm{p} + \mathscr{k}_\mathrm{s})}{h}.
\end{equation}
The branches $\mathscr{k}_\mathrm{p}$ and $\mathscr{k}_\mathrm{s}$ must be chosen consistently with the dispersion diagram in Figure \ref{Fig_1}e, ensuring the real parts vary continuously in frequency.
By choosing positive wavenumbers, ${\mathscr{k}_\mathrm{p} + \mathscr{k}_\mathrm{s} \in \mathbb{N}}$ and can be denoted by $n$ to obtain \eqref{Eq_sum_wavenumbers_ep}.
Additionally, from \eqref{Eq_Bloch_wn_gen} it follows that ${\mathrm{Im}\{ \mathcal{K}_\mathrm{p} \} = \mathrm{Im}\{ \mathcal{K}_\mathrm{s} \}}$ within the gap.
%

The result that ${\gamma_\mathrm{s}=\gamma_\mathrm{p}^\ast}$, where $^{\ast}$ denotes complex conjugation, is now employed to study the behaviour of the eigenvectors within the EP gap.
By definition, ${\tensortwo{T} \vect{V}_\mathrm{p} = \gamma_\mathrm{p} \vect{V}_\mathrm{p}}$, and hence ${\tensortwo{T}^\ast \vect{V}_\mathrm{p}^\ast = \gamma_\mathrm{p}^\ast \vect{V}_\mathrm{p}^\ast}$.
%
%
The transfer matrix satisfies the symmetry relation ${\tensortwo{T}^\ast = \tensortwo{P} \ \tensortwo{T} \ \tensortwo{P}}$, where \\ ${\tensortwo{P} = \mathrm{diag}\left(1, -1, -1, 1 \right)}$.
Consequently, ${\tensortwo{T} \ \tensortwo{P}\vect{V}_\mathrm{p}^\ast=\gamma_\mathrm{p}^\ast\tensortwo{P}\vect{V}_\mathrm{p}^\ast}$, showing that an eigenvector associated with ${\gamma_\mathrm{p}^\ast=\gamma_\mathrm{s}}$ is ${\vect{V}_\mathrm{s}=\tensortwo{P}\vect{V}_\mathrm{p}^\ast}$.
%
%
In terms of the eigenvector components, this becomes
\begin{equation}
    \vect{V}_\mathrm{s} = \begin{pmatrix}
    u_x^\ast \\
    - u_z^\ast \\
    - \sigma_{zz}^\ast \\
    \sigma_{xz}^\ast
    \end{pmatrix} = \begin{pmatrix}
    |u_x| \exp{(-i \varphi_{u_x})} \\
    |u_z| \exp{(i\pi - i\varphi_{u_z})} \\
    |\sigma_{zz}| \exp{(i\pi - i\varphi_{\sigma_{zz}})} \\
    |\sigma_{xz}| \exp{(-i \varphi_{\sigma_{xz}})}
    \end{pmatrix},
\end{equation}
where the phases are defined modulo $2\pi$.
Then, corresponding components of $\vect{V}_\mathrm{p}$ and $\vect{V}_\mathrm{s}$ have equal magnitudes, while their phases satisfy either $\varphi_{v_\mathrm{s}} = - \varphi_{v_\mathrm{p}}$ or $\varphi_{v_\mathrm{s}} = \pi - \varphi_{v_\mathrm{p}}$, depending on the sign of the corresponding entry of $\tensortwo{P}$.

\section{Energy coefficients and wave fields} \label{Append_C}
\setcounter{equation}{0}
\renewcommand{\theequation}{C.\arabic{equation}}

To derive the energy coefficients for the finite layered medium, we determine the wave fields in each layer and in the half-spaces.
When a compressional wave, denoted as $\vect{u}_\mathrm{p}^{\mathrm{in}}$, is incident on the medium from the left half-space (LHS), part of its energy is reflected while the rest is transmitted.
At oblique incidence, mode conversion produces both compressional and shear waves in the reflected and transmitted fields.
Accordingly, each layer contains forward- and backward-propagating compressional and shear waves, while the right half-space (RHS) contains the transmitted compressional and shear waves.
The displacements in the half-spaces and layers are given by
\begin{align}
    \vect{u}_{\textsc{lhs}}  &= \vect{u}_\mathrm{p}^{\mathrm{in}} + \vect{u}_\mathrm{p}^{\mathrm{rf}} + \vect{u}_\mathrm{s}^{\mathrm{rf}}, \quad \vect{u}_{\textsc{rhs}}  = \vect{u}_\mathrm{p}^{\mathrm{tr}} + \vect{u}_\mathrm{s}^{\mathrm{tr}}, \label{Eq_wave_fields_disp_1} \\
    \vect{u}_\ell  & = \vect{u}_{\mathrm{p}_\ell}^{\mathrm{tr}} + \vect{u}_{\mathrm{s}_\ell}^{\mathrm{tr}} + \vect{u}_{\mathrm{p}_\ell}^{\mathrm{rf}} + \vect{u}_{\mathrm{s}_\ell}^{\mathrm{rf}},  \label{Eq_wave_fields_disp_2}
\end{align}
recalling $\ell \in \{1,2\}$ indicates the layers.
%
Here, the superscripts $^\mathrm{tr}$ and $^\mathrm{rf}$ denote transmitted and reflected waves propagating in the positive and negative $z$-directions, respectively.
With $\Bar{u}_{\mathrm{m}_{\ell}}^\mathrm{tr}$ and $\Bar{u}_{\mathrm{m}_{\ell}}^\mathrm{rf}$ indicating the corresponding complex amplitude, each field is given by
\begin{align}
    \vect{u}_\mathrm{p}^{\mathrm{in}}(x,z) &= \Bar{u}_\mathrm{p}^{\mathrm{in}} \vect{P}_{\mathrm{p}}^+(\theta_{\mathrm{in}}) \exp{\left(i \mathcal{Z}^+(k_{\mathrm{p}},\theta_{\mathrm{in}})\right)}, \label{Eq_disp_field_1}\\
    \vect{u}_{\mathrm{m}_\ell}^{\mathrm{tr}}(x,z) &= \Bar{u}_{\mathrm{m}_{\ell}}^\mathrm{tr} \vect{P}_{\mathrm{m}}^+ (\theta_{\mathrm{m}_{\ell}}^\mathrm{tr}) \exp{(i \mathcal{Z}^+ (k_{\mathrm{m}_{\ell}}, \theta_{\mathrm{m}_{\ell}}^\mathrm{tr}))}, \label{Eq_disp_field_2}\\
    \vect{u}_{\mathrm{m}_\ell}^{\mathrm{rf}}(x,z) &= \Bar{u}_{\mathrm{m}_{\ell}}^\mathrm{rf} \vect{P}_{\mathrm{m}}^- (\theta_{\mathrm{m}_{\ell}}^\mathrm{rf}) \exp{(i \mathcal{Z}^- (k_{\mathrm{m}_{\ell}}, \theta_{\mathrm{m}_{\ell}}^\mathrm{rf}))}. \label{Eq_disp_field_3}
\end{align}
%
%
The quantities ${\vect{P}_{\mathrm{p}}^\pm(\theta) = (\sin{(\theta)},\pm \cos{(\theta)})}$ and ${\vect{P}_{\mathrm{s}}^\pm(\theta) = (\mp \cos{(\theta)},\sin{(\theta)})}$ denote the polarisation vectors, and ${\mathcal{Z}^\pm(k,\theta) = k(x \sin{\theta} \pm z \cos{\theta})}$.
Moreover, ${k_{\mathrm{m}_{\ell}} = \omega / c_{\mathrm{m}_{\ell}}}$ is the wavenumber, where $c_{\mathrm{m}_{\ell}}$ indicates the phase speed for mode $\mathrm{m}$ in layer $\ell$.
In general, ${c_{\mathrm{p}} = \sqrt{(\lambda + 2\mu)/\rho}}$ and ${c_{\mathrm{s}} = \sqrt{\mu/\rho}}$.
The Lam\'{e} parameters $\lambda$ and $\mu$ are calculated from $E$ and $\nu$ as ${\lambda = E \nu(1+\nu)^{-1}(1-2\nu)^{-1}}$ and ${\mu = E/(2(1+\nu))}$.
%

By Snell's law, the tangential wavenumber is conserved across the interfaces. 
Consequently, the transmitted and reflected waves of a given mode have the same propagation angle within each layer, so that ${\theta_{\mathrm{m}_\ell}^\mathrm{tr} = \theta_{\mathrm{m}_\ell}^\mathrm{rf} \coloneq \theta_{\mathrm{m}_\ell}}$.
In the half-spaces the reflected and transmitted angles for compressional waves equals the incident angle, i.e., ${\theta_{\mathrm{p}_0}^\mathrm{tr} = \theta_{\mathrm{p}_0}^\mathrm{rf} = \theta_{\mathrm{in}} \coloneq \theta_\mathrm{p}}$, while the corresponding reflected angle for shear waves is denoted by $\theta_\mathrm{s}$ such that ${\theta_{\mathrm{s}_0}^\mathrm{tr} = \theta_{\mathrm{s}_0}^\mathrm{rf} \coloneq \theta_\mathrm{s}}$.
Additionally
\begin{equation} \label{Eq_snells_law} 
    \frac{\sin{\theta_{\mathrm{p}}}}{c_{\mathrm{p}_0}} = \frac{\sin{\theta_{\mathrm{s}}}}{c_{\mathrm{s}_0}} = \frac{\sin{\theta_{\mathrm{p}_1}}}{c_{\mathrm{p}_1}} = \frac{\sin{\theta_{\mathrm{s}_1}}}{c_{\mathrm{s}_1}} = \frac{\sin{\theta_{\mathrm{p}_2}}}{c_{\mathrm{p}_2}} = \frac{\sin{\theta_{\mathrm{s}_2}}}{c_{\mathrm{s}_2}}.
\end{equation}
Here, $c_{\mathrm{p}_0}$ and $c_{\mathrm{s}_0}$ denote the compressional and shear phase speeds in the half-spaces, respectively.

The wave amplitudes are determined by enforcing continuity of displacement and traction at every interface. 
In terms of the state vectors, these are written as
%
%
\begin{alignat*}{2}
    \vect{V}_{h}(x,0) &= \vect{V}_{1}(x,0), \quad \vect{V}_{1}(x,\alpha h + a) &&= \vect{V}_{2}(x,\alpha h + a), \\ 
    \vect{V}_{2}(x,\beta h) &= \vect{V}_{1}(x,\beta h), \quad \quad \vect{V}_{2}(x,\mathscr{n} h) &&= \vect{V}_{h}(x,\mathscr{n} h). 
\end{alignat*}
Here $\mathscr{n}$ is the number of unit cells, while ${\alpha \in \{0,1,\ldots,\mathscr{n}-1}\}$ and ${\beta \in \{1,2,\ldots,\mathscr{n}-1}\}$ index the Layer 1–Layer 2 and Layer 2–Layer 1 interfaces, respectively.
The vectors $\vect{V}_{h}$, $\vect{V}_{1}$, and $\vect{V}_{2}$ denote the state vectors in the half-spaces, Layer 1, and Layer 2, respectively, as defined in \eqref{Eq_transfer_matrix_state}.

Substituting the wave fields into the boundary conditions together with Snell's law yields a linear system for the unknown wave amplitudes.
We define the reflection and transmission coefficients as
\begin{equation} \label{Eq_coeff_hp}
    \mathscr{R}_\mathrm{p} = \frac{\Bar{u}_{\mathrm{p}_0}^\mathrm{rf}}{\Bar{u}_\mathrm{p}^{\mathrm{in}}}, \quad \mathscr{R}_\mathrm{s} = \frac{\Bar{u}_{\mathrm{s}_0}^\mathrm{rf}}{\Bar{u}_\mathrm{p}^{\mathrm{in}}}, \quad \mathscr{T}_\mathrm{p} = \frac{\Bar{u}_{\mathrm{p}_0}^\mathrm{tr}}{\Bar{u}_\mathrm{p}^{\mathrm{in}}}, \quad \mathscr{T}_\mathrm{s} = \frac{\Bar{u}_{\mathrm{s}_0}^\mathrm{tr}}{\Bar{u}_\mathrm{p}^{\mathrm{in}}},
\end{equation}
\begin{equation} \label{Eq_coeff_layers}
    \mathscr{R}_{\mathrm{m}_\ell}^{(\alpha+1)} = \frac{\Bar{u}_{\mathrm{m}_\ell}^\mathrm{rf}}{\Bar{u}_\mathrm{p}^{\mathrm{in}}}, \quad \mathscr{T}_{\mathrm{m}_\ell}^{(\alpha+1)} = \frac{\Bar{u}_{\mathrm{m}_\ell}^\mathrm{tr}}{\Bar{u}_\mathrm{p}^{\mathrm{in}}}.
\end{equation}
The resulting system comprises $4(2 \mathscr{n} + 1)$ equations for the amplitudes in the half-spaces and the individual layers.
Solving this system yields the reflection and transmission coefficients.
To obtain the displacement fields shown in Figures \ref{Fig_3}c-\ref{Fig_3}e, we set $\Bar{u}_\mathrm{p}^{\mathrm{in}} = 1$, determine the remaining amplitudes from the linear system, and substitute them into the wave-field expressions in \eqref{Eq_wave_fields_disp_1}-\eqref{Eq_disp_field_3}.
Importantly, $\mathscr{R}_\mathrm{m}$ denotes the reflection coefficient for mode $\mathrm{m}$ at the interface between the left half-space and the finite layered medium, whereas $\mathscr{T}_\mathrm{m}$ denotes the transmission coefficient through the entire layered structure.

We now calculate the energy coefficients from the time-averaged energy flux.
For time-harmonic elastic waves, the time-averaged intensity vector $\langle \vect{I} \rangle$ is given by \cite{Royer}
\begin{equation} \label{Eq_elasticity_intensity_vector_re}
    \langle \vect{I} \rangle = - \frac{1}{2} \mathrm{Re}\{i \omega \tensortwo{\sigma} \cdot \vect{u}\}.
\end{equation}
The energy flux through an interface is determined by the magnitude of the $z$-component of $\langle \vect{I} \rangle$.
For the wave fields in the half-spaces, the corresponding intensities are
\begin{equation}
    \langle I_\mathrm{p}^{\mathrm{in}} \rangle = \omega^2 |\Bar{u}_\mathrm{p}^{\mathrm{in}}|^2 \kappa_\mathrm{p}, \quad
\end{equation}
\vspace{-3ex}
\begin{alignat}{2}
    \langle I_\mathrm{p}^\mathrm{tr} \rangle &= \omega^2 |\Bar{u}_{\mathrm{p}_0}^\mathrm{tr}|^2 \kappa_\mathrm{p}, \quad
    \langle I_\mathrm{p}^\mathrm{rf} \rangle &&= \omega^2 |\Bar{u}_{\mathrm{p}_0}^\mathrm{rf}|^2 \kappa_\mathrm{p}, \\
    \langle I_\mathrm{s}^\mathrm{tr} \rangle &= \omega^2 |\Bar{u}_{\mathrm{s}_0}^\mathrm{tr}|^2 \kappa_\mathrm{s}, \quad \langle I_\mathrm{s}^\mathrm{rf} \rangle &&= \omega^2 |\Bar{u}_{\mathrm{s}_0}^\mathrm{rf}|^2 \kappa_\mathrm{s},
\end{alignat}
with $\kappa_\mathrm{p} = (\lambda_0 + 2\mu_0) \cos{\theta_\mathrm{p}}/2 c_{\mathrm{p}_0}$ and $\kappa_\mathrm{s} = \mu_0 \cos{\theta_\mathrm{s}}/2 c_{\mathrm{s}_0}$.
The reflected and transmitted energy coefficients are obtained by normalising each reflected or transmitted energy flux by the incident energy flux.
Thus, we obtain
\begin{alignat*}{2}
    \mathcal{E}_\mathrm{p}^\mathrm{rf} &= \frac{\langle I_{\mathrm{p}}^\mathrm{rf} \rangle}{\langle I_\mathrm{p}^{\mathrm{in}} \rangle}  = |\mathscr{R}_\mathrm{p}|^2, \ \mathcal{E}_\mathrm{s}^\mathrm{rf} &&= \frac{\langle I_\mathrm{s}^\mathrm{rf} \rangle}{\langle I_\mathrm{p}^{\mathrm{in}} \rangle} = \frac{c_{\mathrm{s}_0} \cos{\theta_\mathrm{s}}}{c_{\mathrm{p}_0} \cos{\theta_\mathrm{p}}}|\mathscr{R}_\mathrm{s}|^2, \\
    \mathcal{E}_\mathrm{p}^\mathrm{tr} &= \frac{\langle I_\mathrm{p}^{\mathrm{tr}} \rangle}{\langle I_\mathrm{p}^{\mathrm{in}} \rangle}  = |\mathscr{T}_\mathrm{p}|^2, \ \mathcal{E}_\mathrm{s}^\mathrm{tr} &&= \frac{\langle I_\mathrm{s}^\mathrm{tr} \rangle}{\langle I_\mathrm{p}^{\mathrm{in}} \rangle} = \frac{c_{\mathrm{s}_0} \cos{\theta_\mathrm{s}}}{c_{\mathrm{p}_0} \cos{\theta_\mathrm{p}}} |\mathscr{T}_\mathrm{s}|^2,
\end{alignat*}
%
%
Additionally, conservation of energy imposes that ${\mathcal{E}_\mathrm{p}^\mathrm{rf} + \mathcal{E}_\mathrm{s}^\mathrm{rf} + \mathcal{E}_\mathrm{p}^\mathrm{tr} + \mathcal{E}_\mathrm{s}^\mathrm{tr} = 1}$.

\bibliographystyle{unsrt}
\bibliography{biblio}

@book{Kinsler,
   author = {L.~E. Kinsler and A.~R. Frey and A.~B. Coppens and J.~V. Sanders},
   title = {Fundamentals of Acoustics},
   publisher = {John Wiley \& Sons},
   address = {Hoboken},
   edition = {4th},
   year = {2000},
}

@book{Blackstock,
   author = {D.~T. Blackstock},
   title = {Fundamentals of Physical Acoustics},
   publisher = {John Wiley \& Sons},
   address = {New York},
   year = {2000},
}

@book{Anselmet,
   author = {F. Anselmet and P.-O. Mattei},
   title = {Acoustics, Aeroacoustics and Vibrations},
   publisher = {ISTE Ltd},
   address = {London},
   year = {2016}
}

@book{Graf,
   author = {K.F. Graff},
  title={Wave Motion in Elastic Solids},
  year={1991},
  publisher={Dover Publications},
  address={New York},
  note={Originally published: Oxford University Press, 1975}
}

@book{Royer,
   author = {D. Royer and T. Valier-Brasier},
   title = {Elastic waves in solids, volume 1: Propagation},
   publisher = {ISTE Ltd and John Wiley \& Sons},
   address = {London},
   year = {2022}
}

@book{Wei,
   author = {P. Wei},
   title = {Theory of Elastic Waves},
   publisher = {Springer Nature},
   address = {Singapore},
   edition = {1st},
   year = {2023}
}

@book{Whitham,
   author = {G.~B. Whitham},
   title = {Linear and nonlinear waves},
   publisher = {John Wiley \& Sons},
   year = {1999},
   note={Originally published: Wiley, New York, 1974}
}

@book{Brekhovskikh,
   author = {L. Brekhovskikh},
   title = {Waves in layered media},
   publisher = {Academic Press},
   address = {New York, USA},
   edition = {2nd},
   year = {1980}
}

@book{Allard,
   author = {J.~F. Allard and N. Atalla},
   title = {Propagation of Sound in Porous Media: Modelling Sound Absorbing Materials},
   publisher = {John Wiley \& Sons},
   address = {Hoboken},
   year = {2009}
}

@book{Jimenez,
   author = {N. Jim\'{e}nez and O. Umnova and J.-P. Groby},
   title = {Acoustic Waves in Periodic Structures, Metamaterials, and Porous Media: From Fundamentals to Industrial Applications},
   publisher = {Springer Nature},
   address = {Cham},
   year = {2021}
}

@book{Brillouin,
   author = {L. Brillouin},
   title = {Wave Propagation in Periodic Structures: Electric Filters and Crystal Lattices},
   publisher = {Dover Publications},
   address = {New York},
   edition = {2nd},
   year = {1953},
}

@book{Kato,
   author = {T. Kato},
   title = {Perturbation Theory for Linear Operators},
   publisher = {Springer-Verlag},
   address = {Berlin, Heidelberg},
   edition = {2nd},
   year = {1995}
}

@article{Nayfeh91,
   author = {A.~H. Nayfeh},
   title = {The general problem of elastic wave propagation in multilayered anisotropic media},
   journal = {J. Acoust. Soc. Am.},
   volume = {89},
   number = {4},
   pages = {1521-1531},
   year = {1991}
}

@article{Backus62,
   author = {G.~E. Backus},
   title = {Long-wave elastic anisotropy produced by horizontal layering},
   journal = {J. Geophys. Res.},
   volume = {67},
   number = {11},
   pages = {4427-4440},
   year = {1962}
}

@article{Postma55,
   author = {G.~W. Postma},
   title = {Wave propagation in a stratified medium},
   journal = {Geophysics},
   volume = {20},
   number = {4},
   pages = {780-806},
   year = {1955}
}

@article{Folds77,
   author = {D.~L. Folds and C.~D. Loggins},
   title = {Transmission and reflection of ultrasonic waves in layered media},
   journal = {J. Acoust. Soc. Am.},
   volume = {62},
   number = {5},
   pages = {1102-1109},   year = {1977},
}

@article{Scharnhorst83,
   author = {K.~P. Scharnhorst},
   title = {Properties of acoustic and electromagnetic transmission coefficients and transfer matrices of multilayered plates},
   journal = {J. Acoust. Soc. Am.},
   volume = {74},
   number = {6},
   pages = {1883-1886},
   year = {1983}
}

@article{Brouard95,
   author = {B. Brouard and D. Lafarge and J.~F. Allard},
   title = {A general method of modelling sound propagation in layered media},
   journal = {J. Sound Vib.},
   volume = {183},
   number = {1},
   pages = {129-142},
   year = {1995}
}

@article{Thomson50,
   author = {W.~T. Thomson},
   title = {Transmission of Elastic Waves through a Stratified Solid Medium},
   journal = {J. Appl. Phys. },
   volume = {21},
   number = {2},
   pages = {89-93},
   year = {1950}
}

@article{Srikantha11,
   author = {A. Srikantha Phani},
   title = {On elastic waves and related phenomena in lattice materials and structures},
   journal = {AIP Adv.},
   volume = {1},
   number = {4},
   pages = {041602},
   year = {2011}
}

@article{Alizadeh25,
   author = {V. Alizadeh and A.~V. Amirkhizi},
   title = {In-plane linear stress waves in layered media: I. Non-Hermitian degeneracies and modal chirality},
   journal = {Wave Motion},
   volume = {139},
   pages = {103610},
   year = {2025},
   type = {Journal Article}
}

@article{Phani06,
   author = {A. Srikantha Phani and J. Woodhouse and N.~A. Fleck},
   title = {Wave propagation in two-dimensional periodic lattices},
   journal = {J. Acoust. Soc. Am.},
   volume = {119},
   number = {4},
   pages = {1995-2005},
   year = {2006},
}

@article{Mead96,
   author = {D.~J. Mead},
   title = {Wave propagation in continuous periodic structures: research contributions from Southamptom, 1964-1995},
   journal = {J. Sound Vib.},
   volume = {190},
   number = {3},
   pages = {495-524},
   year = {1996},
}

@article{Nunez25,
   author = {G. Núñez and W.~J. Parnell and R.~C. Assier},
   title = {Effective mass density for wave propagation in layered media: a study of the elastic/acoustic transition},
   journal = {Proc. R. Soc. A.},
   volume = {481},
   number = {2318},
   pages = {20250162},
   year = {2025},
}

@article{Hussein09,
   author = {M.~I. Hussein},
   title = {Reduced Bloch mode expansion for periodic media band structure calculations},
   journal = {Proc. R. Soc. A.},
   volume = {465},
   number = {2109},
   pages = {2825-2848},
   year = {2009}
}

@article{Kushwaha93,
   author = {M.~S. Kushwaha and P. Halevi and L. Dobrzynski and B. Djafari-Rouhani},
   title = {Acoustic band structure of periodic elastic composites},
   journal = {Phys. Rev. Lett.},
   volume = {71},
   number = {13},
   pages = {2022-2025},
   year = {1993}
}

@article{Sigalas05,
   author = {M. Sigalas and M.~S. Kushwaha and E.~N. Economou and M. Kafesaki and I.~E. Psarobas and W. Steurer},
   title = {Classical vibrational modes in phononic lattices: theory and experiment},
   journal = {Z. Kristallogr.},
   volume = {220},
   number = {9-10},
   pages = {765-809},
   year = {2005}
}

@article{Kushwaha94,
   author = {M.~S. Kushwaha and P. Halevi and G. Martinez and L. Dobrzynski and B. Djafari-Rouhani},
   title = {Theory of acoustic band structure of periodic elastic composites},
   journal = {Phys. Rev. B, Condens. Matter Mater. Phys.},
   volume = {49},
   number = {4},
   pages = {2313-2322},
   year = {1994}
}

@article{Hussein06,
   author = {M.~I. Hussein and G.~M. Hulbert and R.~A. Scott},
   title = {Dispersive elastodynamics of 1D banded materials and structures: analysis},
   journal = {J. Sound Vib.},
   volume = {289},
   number = {4-5},
   pages = {779-806},
   year = {2006}
}

@article{Lee15,
   author = {H.~J. Lee and J.~K. Lee and Y.~Y. Kim},
   title = {Elastic metamaterial-based impedance-varying phononic bandgap structures for bandpass filters},
   journal = {J. Sound Vib.},
   volume = {353},
   pages = {58-74},
   year = {2015}
}

@article{Esquivel94,
   author = {R. Esquivel-Sirvent and G.~H. Cocoletzi},
   title = {Band structure for the propagation of elastic waves in superlattices},
   journal = {J. Acous. Soc. Am.},
   volume = {95},
   number = {1},
   pages = {86-90},
   year = {1994}
}

@article{Wen05,
   author = {J. Wen and G. Wang and D. Yu and H. Zhao and Y. Liu},
   title = {Theoretical and experimental investigation of flexural wave propagation in straight beams with periodic structures: Application to a vibration isolation structure},
   journal = {J. Appl. Phys.},
   volume = {97},
   number = {11},
   pages = {114907},
   year = {2005}
}

@article{Cao95,
   author = {W. Cao and W. Qi},
   title = {Plane wave propagation in finite 2-2 composites},
   journal = {J. Appl. Phys.},
   volume = {78},
   number = {7},
   pages = {4627-4632},
   year = {1995}
}

@article{Lu18,
   author = {Y. Lu and A. Srivastava},
   title = {Level repulsion and band sorting in phononic crystals},
   journal = {J. Mech. Phys. Solids},
   volume = {111},
   pages = {100-112},
   year = {2018}
}

@article{Seyranian05,
   author = {A.~P. Seyranian and O.~N. Kirillov and A.~A. Mailybaev},
   title = {Coupling of eigenvalues of complex matrices at diabolic and exceptional points},
   journal = {J. Phys. A: Math. Gen.},
   volume = {38},
   number = {8},
   pages = {1723-1740},
   year = {2005}
}

@article{Heiss00,
   author = {W.~D. Heiss},
   title = {Repulsion of resonance states and exceptional points},
   journal = {Phys. Rev. E},
   volume = {61},
   number = {1},
   pages = {929–932},
   year = {2000}
}

@article{Wu04,
   author = {T.-T. Wu and Z.-G. Huang},
   title = {Level repulsions of bulk acoustic waves in composite materials},
   journal = {Phys. Rev. B},
   volume = {70},
   number = {21},
   pages = {214304},
   year = {2004}
}

@article{Dembowski01,
   author = {C. Dembowski and H.~D. Gräf and H.~L. Harney and A. Heine and W.~D. Heiss and H. Rehfeld and A. Richter},
   title = {Experimental Observation of the Topological Structure of Exceptional Points},
   journal = {Phys. Rev. Lett.},
   volume = {86},
   number = {5},
   pages = {787-790},
   year = {2001}
}

@article{Achilleos17,
   author = {V. Achilleos and G. Theocharis and O. Richoux and V. Pagneux},
   title = {Non-Hermitian acoustic metamaterials: Role of exceptional points in sound absorption},
   journal = {Phys. Rev. B},
   volume = {95},
   number = {14},
   pages = {144303},
   year = {2017}
}

@article{Heiss12,
   author = {W.~D. Heiss},
   title = {The physics of exceptional points},
   journal = {J. Phys. A: Math. Theor.},
   volume = {45},
   number = {44},
   pages = {444016},
   year = {2012}
}

@article{Nennig20,
   author = {B. Nennig and E. Perrey-Debain},
   title = {A high order continuation method to locate exceptional points and to compute Puiseux series with applications to acoustic waveguides},
   journal = {J. Comput. Phys.},
   volume = {412},
   pages = {109425},
   year = {2020}
}

@article{Lawrie22,
   author = {J.~B. Lawrie and B. Nennig and E. Perrey-Debain},
   title = {Analytic mode-matching for accurate handling of exceptional points in a lined acoustic waveguide},
   journal = {Proc. R. Soc. A},
   volume = {478},
   number = {2268},
   pages = {20220484},
   year = {2022}
}

@article{Xiong17,
   author = {L. Xiong and B. Nennig and Y. Auregan and W. Bi},
   title = {Sound attenuation optimization using metaporous materials tuned on exceptional points},
   journal = {J. Acoust. Soc. Am.},
   volume = {142},
   number = {4},
   pages = {2288},
   year = {2017}
}

@article{Heiss90,
   author = {W.~D. Heiss and A.~L. Sannino},
   title = {Avoided level crossing and exceptional points},
   journal = {J. Phys. A: Math. Gen.},
   volume = {23},
   pages = {1167–1178},
   year = {1990}
}

@article{Xiong16,
   author = {L. Xiong and W. Bi and Y. Auregan},
   title = {Fano resonance scatterings in waveguides with impedance boundary conditions},
   journal = {J. Acoust. Soc. Am.},
   volume = {139},
   number = {2},
   pages = {764-72},
   year = {2016},
}

@article{Lustig19,
   author = {B. Lustig and G. Elbaz and A. Muhafra and G. Shmuel},
   title = {Anomalous energy transport in laminates with exceptional points},
   journal = {J. Mech. Phys. Solids},
   volume = {133},
   pages = {103719},
   year = {2019}
}

@article{Wang22,
   author = {W. Wang and A.~V. Amirkhizi},
   title = {Exceptional points and scattering of discrete mechanical metamaterials},
   journal = {Eur. Phys. J. Plus},
   volume = {137},
   number = {414},
   year = {2022}
}

@article{Zhu14,
   author = {X. Zhu and H. Ramezani and C. Shi and J. Zhu and X. Zhang},
   title = {PT-Symmetric Acoustics},
   journal = {Phys. Rev. X},
   volume = {4},
   number = {3},
   pages = {031042},
   year = {2014}
}

@article{Christensen16,
   author = {J. Christensen and M. Willatzen and V.~R. Velasco and M.~H. Lu},
   title = {Parity-Time Synthetic Phononic Media},
   journal = {Phys. Rev. Lett.},
   volume = {116},
   number = {20},
   pages = {207601},
   year = {2016}
}

@article{Camley83,
   author = {R.~E. Camley and B. Djafari-Rouhani and L. Dobrzynski and A.~A. Maradudin},
   title = {Transverse elastic waves in periodically layered infinite and semi-infinite media},
   journal = {Phys. Rev. B},
   volume = {27},
   number = {12},
   pages = {7318-7329},
   ISSN = {0163-1829},
   year = {1983}
}

@article{Robert06,
   author = {S. Robert and J.~M. Conoir and H. Franklin},
   title = {Propagation of elastic waves through two-dimensional lattices of cylindrical empty or water-filled inclusions in an aluminum matrix},
   journal = {Ultrasonics},
   volume = {45},
   number = {1-4},
   pages = {178-87},
   year = {2006}
}

@misc{Nunez26,
   author = {G. Núñez and W.~J. Parnell and R.~C. Assier},
   title = {Effective mass density for wave propagation in periodic layered media in the elasto-acoustic transition},
   year = {2026},
   eprint = {2608.07072},
   archivePrefix = {arXiv},
   url = {https://arxiv.org}
}

@article{Gribniak21,
   author = {V. Gribniak and A. Rimkus and L. Plioplys and I. Misiūnaitė and R. Boris and T. Pravilonis},
   title = {Evaluating mechanical efficiency of glass fibres in a polymer profile},
   journal = {Polym. Test.},
   volume = {102},
   ISSN = {01429418},
   year = {2021}
}

@article{Shakir21,
   author = {I. Shakir Abbood and S.~a. Odaa and K.~F. Hasan and M.~A. Jasim},
   title = {Properties evaluation of fiber reinforced polymers and their constituent materials used in structures – A review},
   journal = {Mater. Today: Proc.},
   volume = {43},
   pages = {1003-1008},
   ISSN = {22147853},
   year = {2021},
}

@article{Suputtamongkol22,
   author = {K. Suputtamongkol and K.~J. Anusavice and J.~J. Mecholsky Jr.},
   title = {Cyclic contact fatigue behavior of baria-silicate glass-ceramics as a function of crystal aspect ratio},
   journal = {J. Eur. Ceram. Soc.},
   volume = {42},
   number = {5},
   pages = {2441-2448},
   year = {2022}
}

@article{Mace12,
   author = {B.~R. Mace and E. Manconi},
   title = {Wave motion and dispersion phenomena: veering, locking and strong coupling effects},
   journal = {J. Acoust. Soc. Am.},
   volume = {131},
   number = {2},
   pages = {1015-28},
   year = {2012}
}

@article{Elbaz22,
   author = {G. Elbaz and A. Pick and N. Moiseyev and G. Shmuel},
   title = {Encircling exceptional points of Bloch waves: mode conversion and anomalous scattering},
   journal = {J. Phys. D: Appl. Phys.},
   volume = {55},
   number = {23},
   year = {2022}
}

@article{Fishman24,
   author = {A. Fishman and G. Elbaz and T. Venkatesh Varma and G. Shmuel},
   title = {Third-order exceptional points and frozen modes in planar elastic laminates},
   journal = {J. Mech. Phys. Solids},
   volume = {186},
   pages = {105590},
   year = {2024}
}

@article{Mokhtari20,
   author = {A.~A. Mokhtari and Y. Lu and Q. Zhou and A.~V. Amirkhizi and A. Srivastava},
   title = {Scattering of in-plane elastic waves at metamaterial interfaces},
   journal = {Int. J. Eng. Sci.},
   volume = {150},
   pages = {103278},
   year = {2020}
}

@article{Ji19,
   author = {W.-Q. Ji and Q. Wei and X.-F. Zhu and D.-J. Wu and X.-J. Liu},
   title = {Extraordinary acoustic scattering in a periodic PT-symmetric zero-index metamaterials waveguide},
   journal = {EPL (Europhysics Letters)},
   volume = {125},
   number = {5},
   pages = {58002},
   year = {2019}
}

@article{El-Ganainy18,
   author = {R. El-Ganainy and K.~G. Makris and M. Khajavikhan and Z.~H. Musslimani and S. Rotter and D.~N. Christodoulides},
   title = {Non-Hermitian physics and PT symmetry},
   journal = {Nat. Phys.},
   volume = {14},
   number = {1},
   pages = {11-19},
   year = {2018}
}

@article{Amirkhizi17,
   author = {A.~V. Amirkhizi},
   title = {Homogenization of layered media based on scattering response and field integration},
   journal = {Mech. Mater.},
   volume = {114},
   pages = {76-87},
   year = {2017}
}

\end{document}